# Extending Mandelbox Fractals with Shape Inversions

Gregg Helt

Genomancer, Healdsburg, CA, USA; gregghelt@gmail.com

## Abstract

The Mandelbox is a recently discovered class of escape-time fractals which use a conditional combination of reflection, spherical inversion, scaling, and translation to transform points under iteration. In this paper we introduce a new extension to Mandelbox fractals which replaces spherical inversion with a more generalized shape inversion. We then explore how this technique can be used to generate new fractals in 2D, 3D, and 4D.

## Mandelbox Fractals

The Mandelbox is a class of escape-time fractals that was first discovered by Tom Lowe in 2010 [5]. It was named the Mandelbox both as an homage to the classic Mandelbrot set fractal and due to its overall boxlike shape when visualized, as shown in Figure 1a. The interior can be rich in self-similar fractal detail as well, as shown in Figure 1b. Many modifications to the original algorithm have been developed, almost exclusively by contributors to the FractalForums online community. Although most explorations of Mandelboxes have focused on 3D versions, the algorithm can be applied to any number of dimensions.

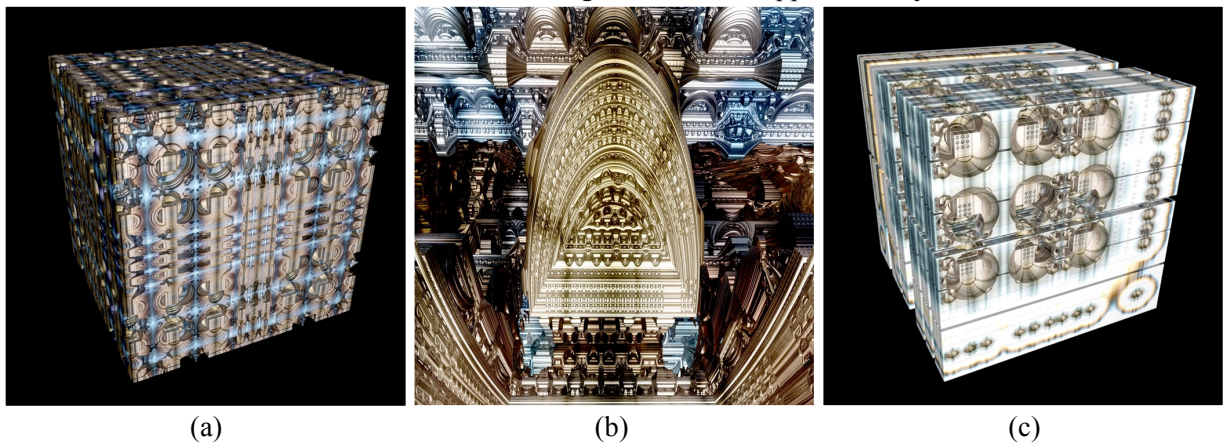

(a)          (b)          (c)

**Figure 1:** *Mandelbox 3D fractal examples: (a) Mandelbox exterior , (b) same Mandelbox, but zoomed in view of small section of interior , (c) Juliabox indexed by same Mandelbox.*

Like the Mandelbrot set and other escape-time fractals, a Mandelbox set contains all the points whose orbits under iterative transformation by a function do not escape. For a basic Mandelbox the function to apply iteratively to each point $P_0$ is defined as a composition of transformations:

$$P_{n+1} = Spherefold_{H,L}(Boxfold_F(P_n)) * S + P_0$$

*Boxfold* and *Spherefold* are modified reflection and spherical inversion transforms, respectively, with parameters $F$, $H$, and $L$ which are described below. $S$ is a scaling parameter. For each initial point $P_0$, if $P_i$ has not exceeded a fixed distance $d$ from the origin after $i$ iterations, $P_0$ is a member of the Mandelbox set $\mathcal{M}(F,H,L,d,i)$. For visualization only points in the set are shown, and in this paper shading of points is determined by a combination of closest approach under orbit to the coordinate axes and the origin.

Similar to how the Mandelbrot set can be considered an index into Julia sets, if we replace translation by $P_0$ (the vector of the point initially iterated on) with translation by a fixed vector $J$ the Mandelbox can be considered as an index into another set of fractals named the Juliabox (see Figure 1c).

## Boxfolds and Spherefolds

For a 3D Mandelbox the *Boxfold* transform consists of conditional reflections across the planes that form the six faces of a cube with side length $2F$ centered on the origin. Using the x-coordinate $P[x]$ of point $P$ as an example, if the coordinate puts the point outside the cube bounds $\pm F$ along the x-axis then the point is reflected through the nearest of the two cube faces that are orthogonal to the x-axis (thus only changing the x-coordinate). Otherwise the x-coordinate is left unchanged. This conditional is repeated for y and z. The overall effect is a folding that moves any points outside the cube closer to the origin.

The *Spherefold* transform is a three-way conditional spherical inversion. If $\|P\|$, the distance from the origin to the point under iteration, is greater than a maximum radius $H$, the point is left unchanged. If $\|P\|$ is less than $H$ but greater than a minimum radius $L$, then the point is transformed by a spherical inversion with $H$ as the inversion radius: $P' = P * H^2 / \|P\|^2$. And if $\|P\|$ is less than $L$ then the point is scaled by a constant, $H^2/L^2$. 2D *Spherefold* inversion and scaling are illustrated in Figure 2b. The overall effect is to move any points within the radius $H$ from the origin further away from the origin.

To better visualize the sequence of transforms in the Mandelbox iteration function, we simplify in Figure 2a by considering a 2D Mandelbox (shown in grayscale in the background), and showing the first iteration on a single point $A$. It is first reflected across the two closest of the four lines $x = \pm F, y = \pm F$ that define the square of the 2D *Boxfold* to yield point $B$, then since $B$ is between the smaller circle (with radius $L$) and the larger circle (with radius $H$) it is inverted with inversion radius $H$ by the 2D *Spherefold* to give point $C$, then (negatively) scaled to give point $D$, and finally offset by the original point $A$ coordinates to give point $E$. For Mandelbox $\mathcal{M}(F,H,L,d,i)$ with $i = 1$ and $d > \|E\|$, $A$ is a member of $\mathcal{M}$.

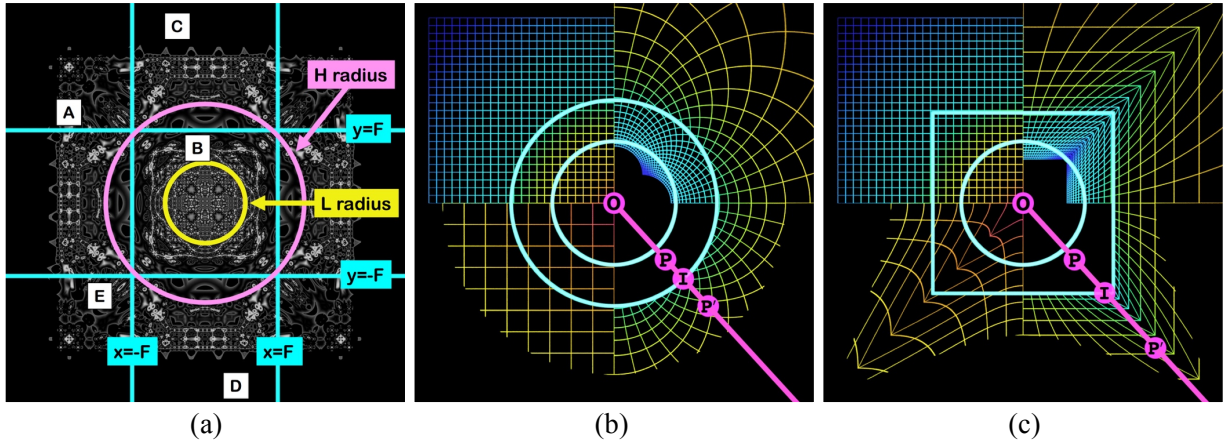

(a)  (b)  (c)

**Figure 2:** *(a) 2D Mandelbox function single iteration of a single point, see text for details , (b) clockwise from upper left: original grid pattern, circle inversion, 2D spherefold conditional inversion, 2D spherefold conditional scaling, (c) same as 2b but with shape inversion in a square.*

## Shape Inversions

Gdaweic [2] introduced a generalization of 2D inversion in a circle to inversion in 2D *star-shaped sets* where circles, ellipses, and regular polygons are all special cases. In [4] we proposed using a simpler subset of star-shaped sets that we call *centered star shapes*, defined as any contiguous shape S with boundary $\partial$ and centroid $O$ in which for all points $Q$ on $\partial$ the line segment $\overline{OQ}$ is contained entirely inside S (or on $\partial$). The shape inversion transformation for any point $P$ with respect to a centered star shape S is identical to that for circle inversion, except that instead of the circle radius $H$ the distance $\|I\|$ from the centroid $O$ to the intersection point $I$ of ray $\overrightarrow{OP}$ with the boundary $\partial$ is used: $P' = P * \|I\|^2 / \|P\|^2$. Circle inversion is thus also a shape inversion where the distance $\|I\|$ for any ray from the origin is a constant, the radius $H$ (see Figure 2b). In comparison Figure 2c shows an example of *Spherefold*-equivalent conditional shape inversion in a square, and illustrates the inversion calculation for a single point $P \Rightarrow P'$.

## 2D Shape Inversion Mandelboxes

We have created new Mandelbox algorithm variants by replacing the circle inversion of radius *H* in the 2D Mandelbox *Spherefold* function with different 2D shape inversions. We have explored various 2D shapes to use for inversion, including the square as shown in Figure 3a, and the Fernandez-Guasti squircle (rounded square, based on Fong parameterization [1]), as shown in Figure 3b. The minimum radius *L* in the *Spherefold* can also be replaced by boundary checking with shapes, for example Figure 3b uses a hexagon. Points within the minimum shape are scaled either by $\|I\|^2 / \|K\|^2$, where K is the intersection point of ray $\overrightarrow{OP}$ with the minimum shape boundary, or alternatively by a constant.

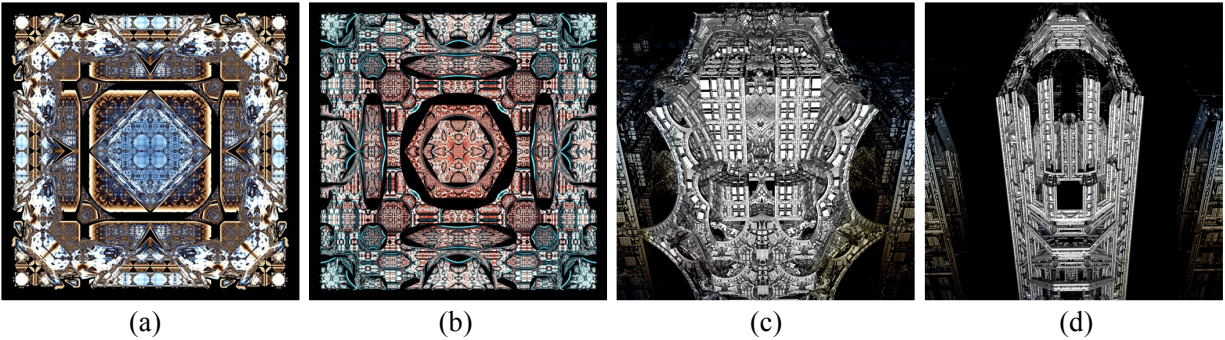

(a)  (b)  (c)  (d)

**Figure 3:** *(a) 2D Mandelbox using square inversion, (b) 2D Mandelbox using FG-squircle inversion and hexagon minimum boundary shape, (c) 3D Mandelbox interior detail with standard spherical inversion, (d) same as 3c but using cube inversion.*

## 3D Shape Inversion Mandelboxes

The only previous work we are aware of on 3D shape inversions is in [6], which focused on ellipsoid inversion. However the definition given above for inversion in centered star shapes is not specific to 2D, thus we can consider 3D and higher dimensional shapes as long as there is a way to determine the intersection point *I* of ray $\overrightarrow{OP}$ with the shape's boundary. We have tried numerous new 3D shape inversions as replacements for spherical inversions in the 3D Mandelbox. Examples are shown in Figure 3d and Figure 4a-d, which for comparison all share the same parameters and interior location as the standard Mandelbox in Figure 3c, other than using a different shape inversion.

We find it striking how much variation the different 3D shape inversions can often introduce locally without disturbing the overall structure. We can add further variety by allowing free rotation of the shapes of inversion, for example compare Figure 3d and Figure 4b, which have identical parameters except for rotation of the cube of inversion. One category of particular interest are shapes that can morph via parameters from spherical to cuboid. Figure 4c shows an example of this where the inversive shape is a rounded cube. In addition we have experimented with inversion shapes that are constructed as the union, intersection or linear blend of more basic shapes, see for example Figure 4d.

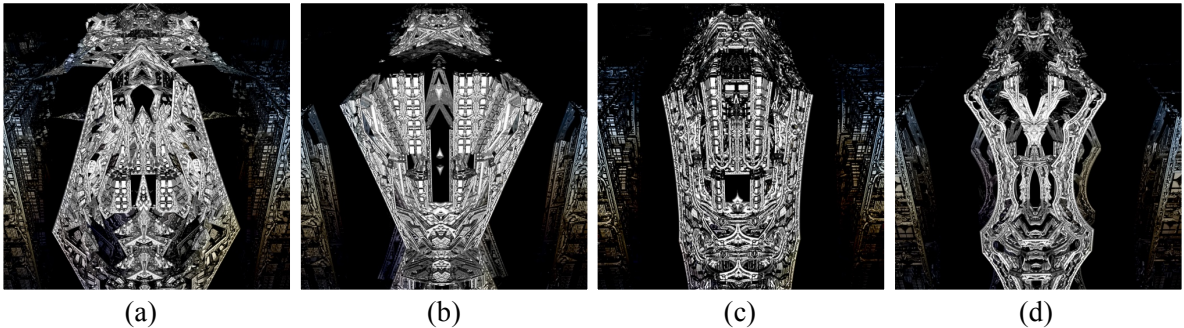

(a)  (b)  (c)  (d)

**Figure 4:** *Same region and Mandelbox as 3c but replacing Spherefold with different shape inversions: (a) octahedron , (b) rotated cube, (c) supershape[3] rounded cube, (d) union of sphere and cube.*

# 4D Shape Inversion Mandelboxes

The standard Mandelbox iterative function extends to higher dimensions, since the *Boxfold* transform generalizes to folding around hypercubes and the *Spherefold* transform generalizes to inversion in hyperspheres. We have explored replacing the 4D Mandelbox *Spherefold* hypersphere inversion with 4D hypercube shape inversion as well as unions, intersections, and linear blends of hyperspheres and hypercubes. We have also tried 4D shape replacements for the 4D *Spherefold* minimum radius. We visualize the results as 3D slices through these new 4D Mandelbox variants as shown in Figure 5.

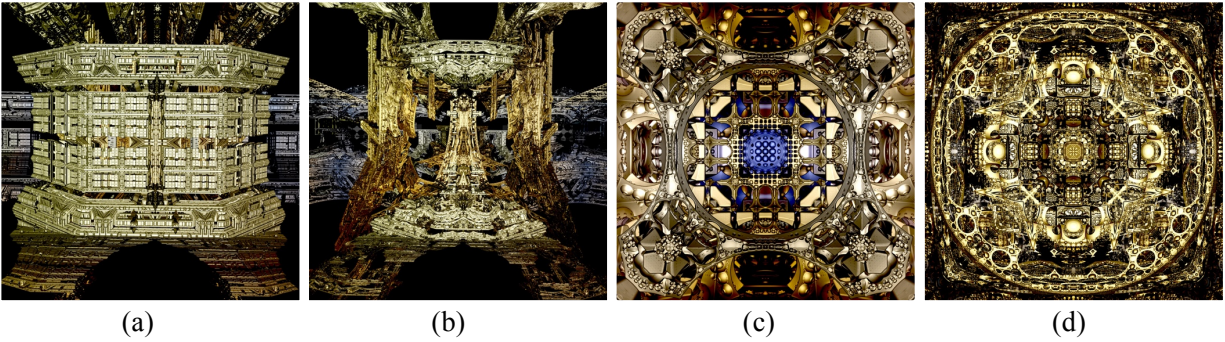

(a)          (b)          (c)          (d)

**Figure 5:** *3D slices through 4D Mandelboxes with 4D shape inversions: (a) interior detail of 4D hypercube inversion, (b) same slice as 5a but using 4D hypersphere/hypercube union inversion, (c) centered view of 4D hypersphere/hypercube intersection inversion, (d) centered view of 4D hypersphere/hypercube blend inversion with 4D hypersphere/hypercube blend minimum shape.*

## Summary

Other modifications to the Mandelbox *Spherefold* transform already exist, but most modify the point distance calculation, and none that we are aware of modify the inversion intersection as is done in shape inversion. We believe this new technique is a significant addition to the artistic toolkit for Mandelboxes and potentially other escape-time fractal algorithms that use spherical inversion. Our code for shape inversion Mandelboxes is open source and available at http://github.com/GreggHelt2/fractalfrags. Additional examples and more information are available at http://www.genomancer.org.

## Acknowledgements

Mandelbox shape inversion algorithms and visualizations were developed as GLSL shader fragments in Fragmentarium, an open source graphics development platform. Additional 2D shape inversion visualizations were developed in JWildfire, an open source algorithmic art program. Many thanks to all who have contributed to Mandelbox development within the FractalForums online community.